\documentclass[floatfix,prb,aps,amsfonts]{revtex4}     
\pdfoutput=1
\usepackage{amssymb,amsmath,graphicx}
\usepackage{verbatim}

\def\xie{ \xi_{\eta}}
\def\xis{ \xi_{\sigma}}

\def\normv{ {\bf e}_{\eta}}
\def\longv{ {\bf e}_{\sigma}}
\def\ter{ \text{t}}
\def\edge{ \text{e}}
\def\ux{ {\bf e}_{x}}
\def\uy{ {\bf e}_{y}}

\def\dv{ {\rm d}}

\begin{document}
\preprint{Submitted to the Journal of Mathematical Physics}

\title{Anisotropic diffusion in continuum relaxation of\\ stepped crystal surfaces}

\author{John Quah and Dionisios Margetis}

\affiliation{Department of Mathematics and Institute for Physical Science
and Technology, University of Maryland, College Park, MD 20742}

\begin{abstract}
We study the continuum limit in 2+1 dimensions of nanoscale anisotropic diffusion processes on crystal surfaces
relaxing to become flat below roughening. Our main result is a continuum law
for the surface flux in terms of a new continuum-scale tensor mobility. The starting point is the Burton, Cabrera and Frank (BCF) theory,
which offers a discrete scheme for atomic steps whose motion drives surface evolution. 
Our derivation is based on the separation of local space variables into fast and slow.
The model includes: (i) {\it anisotropic} diffusion of adsorbed atoms (adatoms) on terraces separating steps; 
(ii) diffusion of atoms along step edges; and (iii) attachment-detachment of atoms at step edges.
We derive a parabolic fourth-order, fully nonlinear partial differential equation (PDE) for the continuum surface height profile.
An ingredient of this PDE is the surface mobility for the adatom flux, which is a nontrivial extension
of the tensor mobility for {\it isotropic} terrace diffusion derived previously by Margetis and Kohn. 
Approximate, separable solutions of the PDE are discussed.
\end{abstract}




\maketitle

%
%
\section{Introduction}
\label{sec:intro}

Theoretical prediction of crystal surface morphological evolution has been an intensively active
area of research for the past several decades. Thanks to advances in computational methods
and experimental techniques, our understanding of the microscopic physics driving 
crystal surface motion continues to improve.~\cite{jeongwilliams99,michelykrug04,evans06} 
Considerable attention has been devoted to nanoscale surface structures evolving via
surface diffusion. Their stability is crucial for their use as building blocks of novel small devices.

Despite continued progress, basic questions on epitaxial phenomena remain unanswered. 
In particular, the relation of microscopic physics to continuum laws, e.g., partial differential equations (PDE's) 
for the surface height profile, is poorly understood. 

Features on crystal surfaces evolve differently according to the
temperature, $T$. Below the roughening temperature, $T_R$, the discrete
nature of the crystal is manifested by macroscopically planar surface
regions (facets) and distinct nanoscale terraces which separate line defects, steps,
of atomic height. The motion of steps drives surface morphological evolution, as first described by Burton, 
Cabrera and Frank (BCF).~\cite{bcf51} 

Continuum theories for crystal surfaces below $T_R$ must be the appropriate limits
of step motion laws and are challenged near facets.~\cite{pimpinellivillain,selkeduxbury95,chameetal96} 
By contrast, above $T_R$ steps are created spontaneously and surfaces appear smooth. In this case, continuum 
laws formulated via thermodynamics and mass conservation are well established.~\cite{pimpinellivillain,mullins,herring51}

Recently, Margetis and Kohn~\cite{margetiskohn06,margetis07} derived systematically the continuum limit in 2+1 dimensions
of a BCF-type model for interacting steps in the absence of material deposition
from above. Their formulation incorporates {\it isotropic} diffusion of adsorbed atoms (adatoms) on terraces and 
atom attachment-detachment at steps; so, the terrace diffusivity is a scalar. 
Their analysis invokes separation of local variables into fast and slow.
A noteworthy element of the resulting theory is the {\it tensor} mobility in Fick's law 
for the adatom flux;~\cite{margetiskohn06,margetis07} the corresponding mobility matrix is diagonal
in the step coordinate system. In this setting, the surface relaxes to become flat via 
an interplay of step energetics and kinetics, and the aspect ratio of step topography brought about
by the tensor character of the mobility.~\cite{margetis07}
Previous continuum theories invoked only a scalar macroscopic mobility, and thus missed the explicit 
influence of topography on evolution;
for a discussion see Ref.~\onlinecite{margetiskohn06}.

In this paper we extend the continuum theory to encompass richer kinetic processes:
{\it anisotropic} adatom diffusion on terraces and atom diffusion {\it along} step edges. In
terrace diffusion, we allow for a non-diagonal diffusivity which explicitly couples adatom fluxes normal
and parallel to step edges. Our goal is to derive continuum laws for surface relaxation that correspond more closely to realistic situations. 
We derive a nonlinear, parabolic fourth-order PDE
for the surface height from a large number of coupled differential equations of step motion. In this PDE, the surface mobility tensor has
off-diagonal elements in the step coordinate system; further, one of the diagonal elements is directly modified by step edge diffusion. 
We find plausible scaling laws with time via approximate, separable PDE solutions.

As a starting point, we adopt the BCF model~\cite{bcf51} by which individual steps move via 
mass conservation for atoms. Each step interacts with its nearest neighbors. Accordingly, coupled differential equations are obtained for step 
positions, which correspond to a discrete scheme. One approach is to solve this scheme numerically.
This approach has been followed mainly for one-dimensional
geometries.~\cite{israelikandel99,israelikandel00,fokthesis06} Another approach is to view the step flow scheme as a discretization
of a continuum evolution equation for the surface height; and {\it derive} this equation in the appropriate limit of small step height
and large number of steps. 
In this paper we focus on the second approach, which lends itself conveniently to 
numerics and prediction of decay laws for
macroscopic surface features in two space dimensions.
 
Most previous continuum approaches to crystal surface morphological
relaxation invoke isotropic physics for each terrace.~\cite{margetiskohn06,rettorivillain88,ozdemirzangwill90,spohn93,shenoy04,margetisetal05} 
However, nanoscale anisotropy is almost ubiquitous, and may stem
from surface reconstruction and the substrate symmetry and structure.~\cite{danker04}

In this paper we focus on terrace diffusion anisotropy,
which is characterized by a tensor diffusivity and can influence pattern formation.~\cite{mongeot00} 
We do not address anisotropy stemming from the step edge orientation dependence of 
parameters such as step line tension and stiffness; the macroscopic limit with such parameters 
is studied in Ref.~\onlinecite{margetiskohn06}. A transformation that relates anisotropic adatom diffusion 
and step edge orientation dependence of step parameters is pointed out in Ref.~\onlinecite{danker04}. This last aspect lies beyond our present scope. 

We also include step edge diffusion~\cite{paulinetal01,pierrelouis01,krug04} for completeness,
since edge diffusion may be important in various experimentally accessible systems.~\cite{jeongwilliams99} In our formalism,
the flux along an edge is driven by variations of the step chemical potential, the change per atom in the step
energy upon addition or removal of atoms at a step edge.
The inclusion of this effect necessarily modifies the surface mobility tensor.

The continuum limit of these processes leads to a generalized relation of the
form ${\bf J}\propto {\bf M}\cdot \nabla\mu$ between
the continuum-scale surface flux, ${\bf J}$, and the continuum step chemical potential, $\mu$.
The coefficient ${\bf M}$ is the macroscopic surface mobility.    
In the curvilinear coordinate system with axes normal and parallel to step edges, ${\bf J}$ is
\begin{equation}\label{eq:mobility}
{\bf J} \propto 
\begin{pmatrix}
M_{11}(|\nabla h|) & M_{12}(|\nabla h|) \\
M_{21}(|\nabla h|) & M_{22}(|\nabla h|)
\end{pmatrix}
\begin{pmatrix}
\partial_{\bot}\mu \\
\partial_{\parallel}\mu
\end{pmatrix}~.
\end{equation}
In this relation, $M_{ij}$ are matrix elements of the tensor mobility $\mathbf{M}$ in the
local representation, $h$ is the surface height profile, and $\partial_\bot$ and $\partial_\parallel$ denote space
derivatives normal and parallel to step edges where the gradient operator is
$\nabla=(\partial_\bot,\partial_\parallel)^T$; cf.~(\ref{eq:updated_flux})--(\ref{eq:M_updated_elements}) 
of Sec.~\ref{sec:slow}.

In previous works that invoke terrace isotropy in 2+1 dimensions,~\cite{margetiskohn06,margetis07} the matrix ${\bf M}$ is
diagonal in the step coordinate system: $M_{12}=M_{21}=0$ with $M_{11}\neq M_{22}$ except in the
special case of diffusion limited kinetics where $M_{11}=M_{22}$.  
This form of mobility does not describe experimental situations where hopping of adatoms
couples the directions normal and parallel to step edges. This coupling is described by setting $D_{12}=D_{21}\neq 0$ in the
diffusivity matrix ${\bf D}$, which in turn yields $M_{12}=M_{21}\neq 0$. Here, we 
determine each $M_{ij}$ explicitly from the step flow model.

There are several critical assumptions inherent to our analysis. 
Our starting model originates from the mesoscale BCF description where steps are replaced by smooth curves.
Hence, we do not consider explicitly atomistic processes which occur at a smaller scale; see e.g. Ref.~\onlinecite{haselwandter}.
In our analysis, the terrace width, a microscopic length, is assumed to be much smaller than: 
(i) the macroscopic length over which the step density varies; (ii) the step radius of curvature; and 
(iii) the length over which the step curvature varies. Step trains 
that satisfy (i)--(iii) are referred to as ``slowly varying''. The terrace width
is comparable to or larger than the step height so that in the continuum limit 
the step density approaches the surface slope. 
We treat monotonic step trains with descending steps and vicinal terraces surrounding a top terrace (peak),
and do not address step motion near a bottom terrace (valley). 

In an attempt to obtain insights into solutions of the derived parabolic PDE
and plausible connections to experiments, we find various 
scaling laws for the continuum-scale height profile, $h$. Here, the term ``scaling law'' describes the time-dependent part $A(t)$ of a separable
solution, $h(\mathbf{r},t)\approx H(\mathbf{r})A(t)$; see Table~\ref{T:scaling laws}. Note that in principle the initial-boundary value
problem for the PDE is not guaranteed to admit separable solutions. This property relies crucially on the 
initial data. Further, nonlinearities of the PDE can play an important role introducing couplings not
captured by scaling scenaria such as ours. We predict scaling laws previously identified 
for isotropic diffusion.~\cite{margetis07}

We do not address the numerical solution of the PDE in this paper.  A promising approach based 
on the finite element method when facets are absent is work in progress.  
Another challenge is to solve the PDE in the presence of facets, where explicit
boundary conditions can be available only from discrete simulations.~\cite{margetisetal06} In the
same vein, the validity of separable PDE solutions is not studied in the present paper.

We assume that the physics of each terrace, although
allowed to be anisotropic, does not vary from one terrace to the next.
Hence, our model cannot fully describe ``surface reconstruction'', the situation
where adatoms on neighboring terraces adapt differently to the missing
bonds at the solid-vapor interface.~\cite{alerhand88,poon90}  We have neglected additional
complications such as sublimation, material
deposition from above, electromigration, and elasticity; the last effect may induce long-range, beyond-nearest-neighbor
step interactions.  The inclusion of these influences in a
more general PDE for the surface height in 2+1 dimensions is the subject of future work.

We organize the remainder of the paper as follows.  In 
Sec.~\ref{sec:background} we present briefly the BCF model; and summarize
a previous derivation~\cite{margetiskohn06,margetis07} of continuum evolution laws 
from discrete equations of step motion for isotropic diffusion.  
In Sec.~\ref{sec:slow} we derive the continuum
limit in the case with anisotropic terrace diffusion and step edge diffusion by placing emphasis
on the relation between surface flux and step chemical potential.  In Sec.~\ref{sec:discussion} we apply
approximately separation of variables to the derived PDE.  Finally, in 
Sec.~\ref{sec:conclusion} we summarize our results and discuss limitations of our theory.

%
%

\section{Background: BCF model and PDE with terrace isotropy}
\label{sec:background}

In this section we review briefly elements of a previous theory~\cite{margetiskohn06,margetis07} that
forms the basis of our analysis. 
The notation, geometry and methodology outlined here serves Sec.~\ref{sec:slow} where we
consider anisotropic terrace diffusion and step edge diffusion. 

We start with the seminal BCF theory,~\cite{bcf51} which introduced a framework to reconcile
the discrete character of crystals in the bulk with the motion of crystal surfaces. 
In this context, crystal surface evolution 
is driven by the motion of steps with atomic height, $a$. 

Motion laws for step edges are determined via mass conservation for atoms:
the step velocity is the sum of fluxes towards and along an edge.
Fluxes result from kinetic processes,
including attachment and detachment of atoms at step edges, 
diffusion of adatoms on terraces, and diffusion of atoms along step edges.
Equilibrium values in kinetic processes are related to step energetics,
namely, the step stiffness and elastic-dipole or entropic step repulsions.~\cite{jeongwilliams99,marchenko80}
We assume that each step interacts only with its nearest neighbors. Beyond-nearest-neighbor
elastic dipole interactions only renormalize the step-step interaction strength and thus are not essentially
different in the continuum limit.~\cite{margetiskohn06}

\subsection{Step geometry}
\label{subsec:geom}

\begin{figure}
\includegraphics[width=3.3in,height=4.0in]{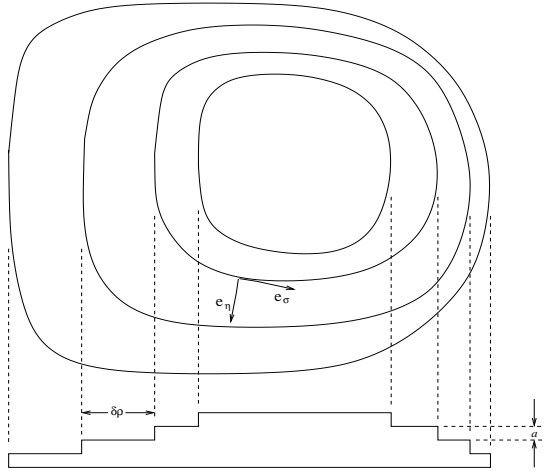}
\caption{Geometry of steps and terraces near surface peak. Top: Projection of step edges to smooth curves on basal plane (top view);
unit vectors $\normv$ and $\longv$ are normal and parallel to step edges.
Bottom: Side view of step train; $a$ is the constant step height 
and $\delta\rho$ is typical terrace width.}\label{F:steptrain}
\end{figure}

In the spirit of BCF,~\cite{bcf51} the edges of steps are projected 
to closed, noncrossing, and non-self-intersecting
smooth curves in a fixed (``basal'') reference plane; see Fig.~\ref{F:steptrain}. These curves
are treated as moving boundaries for the adatom diffusion of each terrace. 

The projection of step edges motivates our choice of
local coordinates.  The steps are descending and are numbered \(i=1,2,\ldots,N\), starting from
the topmost step ($i=1$).  The basal plane position vector
$\mathbf{r}(\eta,\sigma,t)\in\mathbb{R}^{2}$ is a function of time $t$ and
local coordinates \(\eta\) and 
\(\sigma\). The variable $\eta$ identifies the step; $\eta=\eta_i$ for the $i$th step.
The coordinate $\sigma$ indicates the position along an edge, corresponding to 
the angle in polar coordinates; for definiteness, $\sigma$ increases counterclockwise.  
The unit vectors normal and parallel to step edges are \(\normv\) and \(\longv,\) which are
mutually orthogonal and directed toward increasing \(\eta\) and
\(\sigma\).  The associated metric coefficients, which will be needed
below when we compute spatial derivatives, are~\cite{boas84}
\begin{equation}\label{eq:metric coefficients}
\xie := |\partial_{\eta}\mathbf{r}|, \quad \xis :=
|\partial_{\sigma}\mathbf{r}|~.
\end{equation}
The step geometry outlined here remains of course unaltered when we consider terrace anisotropy in Sec.~\ref{sec:slow}.

\subsection{BCF model with step interactions in 2+1 dimensions}
\label{subsec:BCF}

A quantitative discussion of the BCF
theory begins by introducing the adatom density, \(C_{i},\) on the
\(i\)th terrace, $\eta_i<\eta<\eta_{i+1}$.  This \(C_{i}\) satisfies the diffusion equation,
\begin{equation}\label{eq:diffusion equation}
\partial_{t} C_{i} = {\rm div}(\mathbf{D}^{\ter}\cdot \nabla C_{i})~,
\end{equation}
where \(\mathbf{D}^{\ter}\) is a tensor ($2\,$x$\,2$ matrix) diffusivity and $\nabla=(\xie^{-1}\partial_\eta,\xis^{-1}\partial_\sigma)$
is the gradient on the basal plane.
Note that we have omitted from~(\ref{eq:diffusion equation}) terms that describe atom desorption,
electromigration and material deposition from above. A further
simplification emerges from the
``quasisteady approximation'', $\partial_t C_i\approx 0$, which asserts that the time scale for
step motion is much larger than the time scale for terrace diffusion; thus,
the time dependence in $C_i$ enters through the boundary
conditions at step edges.  We define the adatom flux
as \(\mathbf{J}_{i}^{\ter} = -\mathbf{D}^{\ter}\cdot\nabla C_{i}.\)

Robin boundary conditions at the $i$th and $(i+1)$th step edges complement (\ref{eq:diffusion
equation}) to yield a unique solution for $C_i$.  These
conditions emerge from linear kinetics:~\cite{jeongwilliams99,israelikandel99}
\begin{equation}\label{eq:attachment-detachment a}
-J^{\ter}_{i,\bot}(\eta_{i},\sigma,t) = k_{u}[C_{i}(\eta_{i},\sigma,t) -
C_{i}^{eq}(\sigma,t)]~,
\end{equation}
\begin{equation}\label{eq:attachment-detachment b}
J^{\ter}_{i,\bot}(\eta_{i+1},\sigma',t) = k_{d}[C_{i}(\eta_{i+1},\sigma',t) -
C_{i+1}^{eq}(\sigma',t)]~,
\end{equation}
where \(k_{u},k_{d}\) are kinetic rates that account for
the Ehrlich-Schwoebel barrier,~\cite{esI,esII} \(J^{\ter}_{i,\bot}(\eta,\sigma,t) :=
\normv\cdot\mathbf{J}^{\ter}_{i}\) is the transverse component of the adatom flux,
and \(C_{i}^{eq}(\sigma,t)\) is the equilibrium density at the \(i\)th
step edge.

Next, we express \(C_{i}^{eq}\) as a function of step positions by
applying the near-equilibrium thermodynamics law~\cite{jeongwilliams99,israelikandel99}
\begin{equation}\label{eq:Gibbs-Thomson}
C_{i}^{eq}(\sigma) = C_{s}\exp\frac{\mu_{i}(\sigma)}{k_{B}T} \sim
C_{s}\left[1+\frac{\mu_{i}(\sigma)}{k_{B}T}\right]~,
\end{equation}
where \(\mu_{i}\) is the chemical potential of the \(i\)th step. This \(\mu_{i}\) depends on the step edge curvature and
the energy of interactions with other steps.~\cite{jeongwilliams99,israelikandel99,margetiskohn06}  
The linearization in~(\ref{eq:Gibbs-Thomson}) is permissible under typical experimental
conditions.~\cite{tersoff97}

The chemical potential \(\mu_{i}\) can in principle be given as a function of the step curvature and positions.
In Ref.~\onlinecite{margetiskohn06}, $\mu_i$ is found with recourse to differential geometry.
The result reads
\begin{equation}\label{eq:mu_curv}
\mu_i=\frac{\Omega}{a}\left(\frac{1}{\xi_\eta}\partial_{\eta_i} U_i+\kappa_i U_i\right)~,
\end{equation}
where $\Omega$ is the atomic volume, $U_i$ is the total energy per length of the $i$th step edge and $\kappa_i$ is the step edge curvature. We use the definition~\cite{margetiskohn06}
\begin{equation}
U_i=\beta +U_i^{\rm int}~,\label{eq:linet-inter}
\end{equation}
where $\beta$ is the step line tension, assumed here to be a constant, and $U_i^{\rm int}$ is the interaction term which in principle
depends on the positions $\{\eta_j\}$.
For a vicinal surface (i.e., one with sufficiently small slope) and entropic or elastic dipole 
nearest-neighbor interactions, $U_i^{\rm int}$ is~\cite{jeongwilliams99,marchenko80,margetiskohn06,margetis07}
\begin{equation}
U_i^{\rm int}=V_{i,i+1}+V_{i,i-1}~,\label{eq:U-def}
\end{equation}
\begin{equation}
V_{i,i+1}=\frac{g}{3}m_i^2\Phi(\rho_i,\rho_{i+1})~, \qquad \rho_i:=\int_{\eta_0}^{\eta_i}\xi_\eta\,\dv\eta~,\quad m_i:=\frac{a}{\rho_{i+1}-\rho_i}~,
\label{eq:Vi-def}
\end{equation}
where $g$ is a positive constant ($g>0$), $\rho_i$ corresponds to distance in polar coordinates,
$m_i$ is the discrete step density, and $\Phi$ is a shape factor; note that $\Phi(\rho_i,\rho_i)=$const.~\cite{margetiskohn06}

An important remark is in order. Because $C_i^{eq}$ and $\mu_i$ 
are defined as independent of the kinetic processes, the formulation 
for the step chemical potential here carries through unaltered when we introduce
anisotropic terrace diffusion in Sec.~\ref{sec:slow}. 

Lastly, we introduce the step velocity law. By including
diffusion of atoms along the step edge 
with constant edge diffusivity $D^{\edge}$, the normal velocity of the
$i$th step edge is~\cite{danker04,paulinetal01,pierrelouis01,krug04} 
\begin{equation}\label{eq:step velocity}
v_{i} = \normv\cdot\frac{\dv \mathbf{r}_{i}}{\dv t} =
\frac{\Omega}{a}(J^{\ter}_{i-1,\bot} - J^{\ter}_{i,\bot})+
a\partial_{s}\biggl(D^{\edge}\partial_{s}
\frac{\mu_{i}}{k_{B}T}\biggr)~,
\end{equation}
where $\partial_s$ is the space derivative along a step edge;
$\partial_s=\xis^{-1}\partial_\sigma$. The first term in~(\ref{eq:step velocity}) is the contribution of terrace adatom fluxes.
The second term is due to step edge diffusion and
stems from the variation of the step chemical potential, $\mu_i$.
A reasoning for using $\mu_i$ both in edge diffusion and in $C_i^{eq}$ relies on the
fact that $\mu_i$ controls the equilibrium shape of a step. This
equilibrium state is expected to be independent of the kinetic pathway (edge diffusion or attachment-detachment).
So, if mass exchange with the terrace is turned off and relaxation occurs via edge diffusion,
the step attains the same shape as in the case where edge diffusion is turned off and relaxation is allowed only by
attachment-detachment kinetics. This property implies that the thermodynamic driving force 
has to be the same chemical potential, $\mu_i$, in both cases.~\cite{krug-priv}

Equations~(\ref{eq:diffusion equation})--(\ref{eq:step velocity}) in principle lead to a system of coupled
differential equations for the step positions.  This system is a discrete scheme of step flow and has been
solved numerically for straight and circular interacting steps.~\cite{israelikandel99,israelikandel00,fokthesis06}
In this section we focus on (2+1)-dimensional settings with $D^{\edge}=0$.

\subsection{Approximations for slowly varying step train}
\label{subsec:approx}

Evidently, the adatom flux ${\bf J}^{\ter}_i$ plays a pivotal role in
connecting the step velocity to the step chemical potential.  Next, we
find an explicit formula for this flux by solving the diffusion
equation~(\ref{eq:diffusion equation}) approximately following Ref.~\onlinecite{margetiskohn06}. 

The key idea is to consider slowly varying step trains and treat
the local variables $\eta$ and $\sigma$ as fast and slow, respectively.
This assumption enables us to neglect the \(\sigma\) derivatives
in~(\ref{eq:diffusion equation}). Accordingly, for constant ${\bf D}^{\ter}$ the diffusion 
equation for $C_i$ reduces to 
\begin{equation}
\partial_\eta\biggl(\frac{\xis}{\xie}\partial_\eta C_i\biggr)\approx 0~,
\label{eq:Ci-slow}
\end{equation}
which has the explicit solution
\begin{equation}
C_i\approx A_i(\sigma,t)\int_{\eta_i}^\eta \frac{\xie}{\xis}\ \dv\eta^\prime+B_i(\sigma,t)\qquad \eta_i<\eta <\eta_{i+1}~,
\label{eq:Ci-slow-soln}
\end{equation}
where $A_i$ and $B_i$ are integration constants to be determined via the 
boundary conditions~(\ref{eq:attachment-detachment a}),~(\ref{eq:attachment-detachment b}).

For isotropic adatom diffusion~\cite{margetiskohn06} with (scalar) diffusivity $D^{\ter}$ the vector-valued adatom flux is computed by
\begin{equation}
{\bf J}^{\ter}_i=-D^{\ter}\nabla C_i~.
\label{eq:Jter-D}
\end{equation}
By use of~(\ref{eq:attachment-detachment a}) and~(\ref{eq:attachment-detachment b}),
the flux components restricted at $\eta=\eta_i$ are
\begin{align}
\label{eq:discrete normal flux}
J^{\ter}_{i,\bot} &= -\frac{D^{\ter}C_{s}}{k_{B}T}\,\frac{1}{\xis|_i} \frac{\mu_{i+1} -
\mu_{i}}{D^{\ter}\left( \frac{1}{k_{u}\xis|_{i}} +
\frac{1}{k_{d}\xis|_{i+1}} \right) + \int_{\eta_{i}}^{\eta_{i+1}}
\frac{\xie}{\xis} \dv\eta}, \\
\label{eq:discrete parallel flux}
J^{\ter}_{i,\parallel} &= -\frac{D^{\ter}}{\xis|_{i}} \partial_{\sigma} \left\{
\frac{D^{\ter} \left( \frac{C_{i+1}^{eq}}{k_{u}\xis|_{i}} +
\frac{C_{i}^{eq}}{k_{d}\xis|_{i+1}} \right) + C_{i}^{eq}
\int_{\eta_{i}}^{\eta_{i+1}} \frac{\xie}{\xis} \dv\eta}{D^{\ter}\left(
\frac{1}{k_{d}\xis|_{i+1}} + \frac{1}{k_{u}\xis|_{i}} \right) +
\int_{\eta_{i}}^{\eta_{i+1}} \frac{\xie}{\xis} \dv\eta} \right\},
\end{align}
where \(J^{\ter}_{i,\parallel} := \longv\cdot\mathbf{J}^{\ter}_{i}.\)
For details on the anisotropic case see~Sec.~\ref{sec:slow}.

We pause here to review the assumptions underlying the above approximations.
The derivative \(\partial_{\sigma}\) is treated as
\(O(\text{``}\epsilon\text{''})\) in comparison to the derivative
\(\partial_{\eta},\) which is treated as \(O(1)\); $\epsilon\ll 1$. It is reasonable to think of
\(\epsilon\) as being of the order of $a\kappa$ where $\kappa=O(\lambda^{-1})$ is a typical
step curvature and $\lambda$ is a suitable macroscopic length.~\cite{margetiskohn06}
Once the continuum-scale surface flux is derived, the assumptions for the
$\eta$ and $\sigma$ derivatives are relaxed: both derivatives are allowed to be $O(1)$.
An alternative yet equivalent approach based on Taylor expansions at adjacent step edges
is described in Ref.~\onlinecite{margetis07} and in Sec.~\ref{sec:slow} below. 

\subsection{Continuum theory with isotropic diffusion in 2+1 dimensions}
\label{subsec:old continuum}

Step motion laws are viewed as the result of discretizing a PDE for the
continuum-scale surface height profile. In this section we review the continuum limit
of the discrete model~(\ref{eq:diffusion equation})--(\ref{eq:step velocity})
when the physics of each terrace is isotropic (${\bf D}^{\ter}=D^{\ter}$: scalar)
and there is no step edge diffusion ($D^{\edge}=0$).~\cite{margetiskohn06} Accordingly,
we derive a nonlinear fourth-order PDE for the surface height.

First, we summarize the main assumptions applied in
Ref.~\onlinecite{margetiskohn06}. The continuum limit corresponds formally
to taking $a/\lambda\to 0$ where $\lambda$ is a macroscopic length.
The metric coefficients \(\xis\) and \(\xie\) are \(O(\lambda),\) while
the terrace width \(\delta\rho_{i}\) is \(O(a).\)  Therefore, we have
\(
\delta\eta_{i} = \eta_{i+1}-\eta_{i} \sim \delta\rho_{i}\xie^{-1} =
O(a/\lambda) \rightarrow 0.
\)
In this limit, we must keep as fixed, \(O(1)\) quantities the step
density \(m_{i}=a/\delta\rho_{i}\) and the kinetic parameters
\(D^{\ter}/(k_la)\) where $l=u$ or $d$.  

The limiting procedure relies on identifying any discrete variable
$Q_i$ at a step edge ($\eta=\eta_i$) with the interpolation of a continuous, 
sufficiently differentiable function $\widetilde Q(\eta=\eta_i)$.
Thus, $Q_{i+1}-Q_i\approx (\delta\eta_i)\,\partial_\eta\widetilde Q|_i$ where $\mathcal A|_i$ denotes $\mathcal A(\eta_i)$ throughout. 
The following assertions of Ref.~\onlinecite{margetiskohn06} carry through for
the continuum limit of Sec.~\ref{sec:slow}. (i) The step density
approaches the surface slope,
\(
m_{i}\rightarrow m = |\nabla h||_{i} = O(1).
\)
(ii) The unit vector normal to the \(i\)th step edge becomes
\(
\normv|_{i} \rightarrow \normv = -\frac{\nabla h}{|\nabla h|}.
\)
(iii) The step curvature, \(\kappa_{i} = \nabla\cdot\normv|_{i},\)
approaches
\(
\kappa_{i} \rightarrow \kappa = -\nabla\cdot\left( \frac{\nabla
h}{|\nabla h|} \right).
\)
(iv) The step normal velocity, \(v_{i} =
\normv\cdot\dv\mathbf{r}_{i}/\dv t,\) becomes
\(
v_{i}\rightarrow v(\mathbf{r},t) = \frac{\partial_{t}h}{|\nabla h|},
\)
the velocity of the level set with height \(h.\)

\subsubsection{Adatom flux}
\label{ssec:flux}

Next, we outline the continuum limit of the flux components~(\ref{eq:discrete normal
flux}) and (\ref{eq:discrete parallel flux}). The terms on the
right-hand sides of these equations are replaced by series expansions as \(\delta\eta_{i}\rightarrow 0.\)

The resulting continuum limit has the form of a matrix equation
involving the adatom mobility \(\mathbf{M}^{\ter}\), viz.,~\cite{comment}
\begin{equation}\label{eq:continuum flux-potential simple}
\mathbf{J}^{\ter}_i|_i\rightarrow {\bf J}^{\ter}({\bf r},t)=\left( \begin{array}{cc}
J^{\ter}_{\bot} \\
J^{\ter}_{\parallel} \end{array} \right) = -C_{s}\mathbf{M}^{\ter}
\cdot\left( \begin{array}{c}
\partial_{\bot}\mu \\
\partial_{\parallel}\mu \end{array} \right)~, 
\end{equation}
where
\begin{equation}\label{eq:mobility 0}
\mathbf{M}^{\ter} = \frac{D^{\ter}}{k_{B}T}\left(
\begin{array}{cc}
\frac{\displaystyle 1}{\displaystyle 1+q|\nabla h|} & 0 \\
0 & 1 \end{array} \right)~,
\end{equation}
\(\partial_{\bot} = \xie^{-1}\partial_{\eta},\)
\(\partial_{\parallel} = \xis^{-1}\partial_{\sigma}\) and the kinetic
parameter \(q\) is defined by
\begin{equation}\label{eq:kinetic parameter}
q := \frac{2D^{\ter}}{ka}, \qquad k^{-1} := (k_{u}^{-1} + k_{d}^{-1})/2.
\end{equation}
Equation~(\ref{eq:continuum flux-potential simple}) is complemented
with a mass conservation statement for the height profile $h$ and a continuum law for the continuum-scale step chemical
potential $\mu$.

\subsubsection{Continuum step chemical potential}
\label{subsec:continuum step chemical potential}

Next, we invoke~(\ref{eq:mu_curv})--(\ref{eq:Vi-def}) for the step chemical potential $\mu_i$.
Note that we can treat the step edge energy per unit length \(U_{i}\) as the
restriction to \(\eta_{i}\) of a continuous function \(\widetilde U(\eta)\).~\cite{margetiskohn06}  It follows that
$\mu_{i}(\sigma,t) = \frac{\Omega}{a}{\rm div}(\widetilde U\normv)|_i$.

The continuum step chemical potential $\mu({\bf r},t)$ is found by taking the continuum
limit of~(\ref{eq:mu_curv})--(\ref{eq:Vi-def}). The result is~\cite{margetiskohn06,comment} 
\begin{equation}\label{eq:continuum mu}
\mu_i(t)\rightarrow \mu = -\frac{\Omega}{a}\ {\rm div}\Biggl[(\beta + \tilde{g}|\nabla
h|^{2})\frac{\nabla h}{|\nabla h|}\Biggr]~,\qquad \tilde g:=g\Phi(\rho_i,\rho_i)={\rm const.}
\end{equation}
Note that the definition of $\mu_i$ and thus the limit~(\ref{eq:continuum mu}) is not affected by the kinetics;
thus, (\ref{eq:continuum mu}) remains unaltered by the inclusion of step edge diffusion 
and terrace diffusion anisotropy.

\subsubsection{Mass conservation for adatoms}
\label{sssec:mass-conserv}

For $D^{\edge}=0$ the step velocity law~(\ref{eq:step velocity}) 
reduces to the usual mass conservation statement for adatoms.~\cite{margetiskohn06}
Indeed, in the continuum limit the step velocity $v_i$ approaches $\partial_t h/|\nabla h|$.
On the other hand, \(J^{\ter}_{i-1,\bot}|_{i}\) in the
term \(J^{\ter}_{i-1,\bot}|_{i} - J^{\ter}_{i,\bot}|_{i}\) of (\ref{eq:step
velocity}) is replaced by an expression involving \(\mathbf{J}^{\ter}_{i-1}\)
evaluated at \(\eta=\eta_{i-1}\) through  integration of \({\rm div}\mathbf{J}^{\ter}_{i-1}=0\) on the
\((i-1)\)th terrace. This substitution yields a sum that is
recognized as a divergence in the continuum limit:  the right-hand side of
(\ref{eq:step velocity}) approaches
\(-\frac{\Omega}{|\nabla h|} \nabla\cdot\mathbf{J}^{\ter}\) when $D^{\edge}=0$.~\cite{margetiskohn06} The resulting equation is
\begin{equation}\label{eq:mass-cons}
\partial_t h+\Omega\,{\rm div}{\bf J}^{\ter}=0~.
\end{equation}

\subsubsection{Evolution equation for surface height}
\label{sssec:h-evolution}

A PDE for the surface height $h({\bf r},t)$
is found by combination of~(\ref{eq:continuum flux-potential simple}), (\ref{eq:continuum mu}) and~(\ref{eq:mass-cons}):~\cite{margetiskohn06,margetis07}
\begin{equation}
\partial_t h=-B\,{\rm div}\,\biggl\{{\bf \Lambda}^{\ter}\cdot
\nabla\biggl[{\rm div}\biggl(\frac{\nabla h}{|\nabla h|}
+\frac{g_3}{g_1}\,|\nabla h|\nabla h\biggr)\biggr]\biggr\}~,
\label{eq:PDE-2D}
\end{equation}
where
\begin{equation}\label{eq:parameters}
{\bf \Lambda}^{\ter}:=\frac{k_BT}{D^{\ter}}\,{\bf M}^{\ter}~, \qquad g_1:=\beta/a,\quad g_3:=\tilde g/a,
\quad B:=\frac{D^{\ter}C_sg_1\Omega^2}{k_BT}~.
\end{equation}
Evidently, the material parameter $B$ has dimensions (length)$^4$/time and ${\bf \Lambda}^{\ter}$ is dimensionless.

%
\section{Anisotropic diffusion}
\label{sec:slow}

In this section we extend the theory of Sec.~\ref{sec:background} to
cases with a tensor-valued terrace diffusivity \(\mathbf{D}^{\ter}\) and 
a nonzero edge diffusivity $D^{\edge}$, which
offer a more realistic description of diffusion processes on terraces and steps.
Our goal is to derive a PDE for the surface height. A main ingredient is the
surface mobility, which is an extension of~(\ref{eq:mobility 0}).

The terrace diffusivity \(\mathbf{D}^{\ter}\) is assumed to have the
tensor form \(\mathbf{D}^{\ter} = D_{11}\normv\normv + D_{12}\normv\longv
+ D_{21}\longv\normv + D_{22}\longv\longv\). For the sake of some generality,
we do not enforce the symmetry relation $D_{12}=D_{21}$, although this equality is often dictated 
on physical grounds. The components of the surface flux \(\mathbf{J}^{\ter}_{i}\) are related to both
spatial derivatives of the adatom density $C_i$ through the linear relation
\begin{equation}\label{eq:density-flux}
\left( \begin{array}{c}
J^{\ter}_{i,\bot} \\
J^{\ter}_{i,\parallel} \end{array} \right) = -\left( \begin{array}{cc}
D_{11} & D_{12} \\
D_{21} & D_{22} \end{array} \right)\cdot\left( \begin{array}{c}
\xie^{-1} \partial_{\eta} C_{i} \\
\xis^{-1} \partial_{\sigma} C_{i} \end{array}
\right)\qquad \eta_i<\eta <\eta_{i+1}~,
\end{equation}
assuming that no drift term is present, which would arise from an electromigration current.

\subsection{Approximations for fast and slow step variables}
\label{subsec:density-flux}

In this subsection we provide relations for the adatom flux components at step edges
for slowly varying step trains. The starting point is the diffusion equation (\ref{eq:diffusion equation}), which becomes
\begin{equation}\label{eq:tensor diffusion equation}
\frac{\partial}{\partial\eta} \left(\frac{\xis D_{11}}{\xie}
\frac{\partial C_{i}}{\partial\eta}\right) +
\frac{\partial}{\partial\eta} \left(D_{12}
\frac{\partial C_{i}}{\partial\sigma}\right) 
+ \frac{\partial}{\partial\sigma}
\left(D_{21} \frac{\partial C_{i}}{\partial\eta}\right) +
\frac{\partial}{\partial\sigma} \left( \frac{\xie D_{22}}{\xis}
\frac{\partial C_{i}}{\partial\sigma}\right)=0 \qquad \eta_i <\eta <\eta_{i+1}~.
\end{equation}
In particular, for slowly varying step train we invoke the separation of the variables ($\eta,\sigma$) into fast
and slow as outlined in Sec.~\ref{subsec:old continuum}.
Hence, (\ref{eq:tensor diffusion equation}) reduces to~(\ref{eq:Ci-slow}), which is
solved by~(\ref{eq:Ci-slow-soln}). By~(\ref{eq:density-flux}), the corresponding flux components are 
\begin{align}
J^{\ter}_{i,\bot} &\approx -\frac{D_{11}}{\xis} A_{i}(\sigma,t) -
\frac{D_{12}}{\xis} \partial_{\sigma} \left[ B_{i}(\sigma,t) +
A_{i}(\sigma,t) \int_{\eta_{i}}^{\eta} \frac{\xie}{\xis} \dv \eta'
\right]~,\label{eq:anis-flux a} \\
J^{\ter}_{i,\parallel} &\approx -\frac{D_{21}}{\xis} A_{i}(\sigma,t) -
\frac{D_{22}}{\xis}
\partial_{\sigma} \left[ B_{i}(\sigma,t) + A_{i}(\sigma,t)
\int_{\eta_{i}}^{\eta}
\frac{\xie}{\xis} \dv \eta'\right]~.\label{eq:anis-flux b}
\end{align}

Equations~(\ref{eq:anis-flux a}) and~(\ref{eq:anis-flux b}) are simplified when we 
evaluate \(\mathbf{J}^{\ter}_{i}\) at \(\eta = \eta_{i}\). The resulting matrix equation is  
\begin{equation}\label{eq:flux-coefficient relation}
-\xis|_{i} \left( \begin{array}{c}
J^{\ter}_{i,\bot}|_{i} \\
J^{\ter}_{i,\parallel}|_{i} \end{array} \right) = \left( \begin{array}{cc}
D_{11} & D_{12} \\
D_{21} & D_{22} \end{array} \right) \left( \begin{array}{c}
A_{i} \\
\partial_{\sigma} B_{i} \end{array} \right)~.
\end{equation}
By inspection of (\ref{eq:flux-coefficient relation}), the term
\(\partial_{\sigma}B_{i}\) must be treated on equal footing with
\(A_{i},\) since both terms make comparable contributions to the
surface flux.  We proceed to invert the
matrix equation (\ref{eq:flux-coefficient relation}), viewing
\(A_{i}\) and \(\partial_{\sigma}B_{i}\) as integration constants that
we have to eliminate from the boundary conditions
(\ref{eq:attachment-detachment a}) and (\ref{eq:attachment-detachment b}). Thus, we obtain the formula
\begin{equation}\label{eq:coefficient-flux relation}
\left( \begin{array}{c}
A_{i} \\
\partial_{\sigma}B_{i} \end{array} \right) =
-\frac{\xis|_{i}}{|\mathbf{D}^{\ter}|} \left( \begin{array}{cc}
\phantom{-}D_{22} \qquad & -D_{12} \\
-D_{21} \qquad & \phantom{-}D_{11} \end{array} \right) \left( \begin{array}{c}
J^{\ter}_{i,\bot}|_{i} \\
J^{\ter}_{i,\parallel}|_{i} \end{array} \right)~,\qquad |{\bf D}^{\ter}|:=D_{11}D_{22}-D_{12}D_{21}~.
\end{equation}
Note that \(|\mathbf{D}^{\ter}|\) denotes the determinant of \(\mathbf{D}^{\ter}.\)

Next, we apply the boundary conditions (\ref{eq:attachment-detachment
a}) and (\ref{eq:attachment-detachment b}) for atom
attachment-detachment at step edges.  By substituting the solution for
the adatom density $C_i$ into these conditions, we find the relations
\begin{align}
\label{eq:attachment-detachment up}
-J^{\ter}_{i,\bot}(\eta_{i},\sigma,t) &= k_{u} [B_{i}(\sigma,t) -
C_{i}^{eq}(\sigma,t)] \\
\label{eq:attachment-detachment down}
J^{\ter}_{i,\parallel}(\eta_{i+1},\sigma',t) &= k_{d} \left[B_{i}(\sigma',t) +
A_{i}(\sigma',t)\int_{\eta_{i}}^{\eta_{i+1}} \frac{\xie}{\xis}\ \dv \eta
- C_{i+1}^{eq}(\sigma',t)\right].
\end{align}

We eliminate \(B_{i}\) by setting \(\sigma'=\sigma\) in equation
(\ref{eq:attachment-detachment down}), multiplying
(\ref{eq:attachment-detachment up}) by \(k_{d}/k_{u}\) and subtracting
the resulting equation from (\ref{eq:attachment-detachment down}).
Substituting for \(A_{i}\) from (\ref{eq:coefficient-flux
relation}), we arrive at the first desired relation between the surface flux components:
\begin{equation}\label{eq:flux-density 1}
\left( \frac{1}{k_{u}} + \frac{\xis|_{i}D_{22}}{|\mathbf{D}^{\ter}|}
\int_{\eta_{i}}^{\eta_{i+1}} \frac{\xie}{\xis} \dv \eta \right)
J^{\ter}_{i,\bot}|_{i} + \frac{1}{k_{d}} J^{\ter}_{i,\bot}|_{i+1} 
- \frac{\xis|_{i}D_{12}}{|\mathbf{D}^{\ter}|}
\left( \int_{\eta_{i}}^{\eta_{i+1}} \frac{\xie}{\xis} \dv \eta \right)
J^{\ter}_{i,\parallel}|_{i} = C_{i}^{eq} - C_{i+1}^{eq}~.
\end{equation}

We obtain a second relation by exploiting variations in \(\sigma\), which can
be taken to be arbitrarily small; in contrast, changes in $\eta$ are restricted
by $a$ and the requirement of finite slope.  Therefore, we differentiate
(\ref{eq:attachment-detachment up}) with respect to \(\sigma\) and
substitute for \(\partial_{\sigma}B_{i}\) from
(\ref{eq:coefficient-flux relation}).  Subsequently, we neglect \(\partial_{\sigma}
J^{\ter}_{i,\bot}\), consistent with the hypothesis of slowly varying step edge curvature.  
Thus, the second desired relation of the flux components reads
\[
\frac{\xis|_{i}}{|\mathbf{D}^{\ter}|} (D_{21}J^{\ter}_{i,\bot}|_{i} -
D_{11}J^{\ter}_{i,\parallel}|_{i}) - \partial_{\sigma} C_{i}^{eq}=0~,
\]
which in turn becomes
\begin{equation}
\label{eq:flux-density 2}
D_{21}J^{\ter}_{i,\bot}|_{i} - D_{11}J^{\ter}_{i,\parallel}|_{i} =
\frac{C_{s}|\mathbf{D}^{\ter}|}{\xis|_{i}}
\frac{\partial_{\sigma} \mu_{i}}{k_{B}T} =
\frac{C_{s}|\mathbf{D}^{\ter}|}{k_{B}T}\partial_{\parallel}\mu_{i}~.
\end{equation}
Equations (\ref{eq:flux-density 1}) and (\ref{eq:flux-density 2}) suffice for the purpose of taking the continuum limit.

\subsection{Continuum-scale adatom flux}
\label{subsec:limitES}

In this subsection we derive the analogue of (\ref{eq:continuum
flux-potential simple}) and~(\ref{eq:mobility 0}), the relation between continuum adatom
flux and step chemical potential. The resulting terrace mobility, ${\bf M}^{\ter}$, will still need modification
to account for step edge diffusion. 

First, we simplify relations~(\ref{eq:flux-density 1}) and~(\ref{eq:flux-density 2})
for ${\bf J}^{\ter}_i$. Considering \(\delta\eta_{i} = \eta_{i+1}-\eta_{i}\) as small,
we make the approximations
\begin{align*}
\frac{1}{k_{u}} J^{\ter}_{i,\bot}|_{i} + \frac{1}{k_{d}}
J^{\ter}_{i,\bot}|_{i+1} &= \left( \frac{1}{k_{u}} + \frac{1}{k_{d}}
\right)J^{\ter}_{i,\bot}|_{i}\big[1+O(\delta\eta_{i})\big]~, \\
\int_{\eta_{i}}^{\eta_{i+1}} \frac{\xie}{\xis} \dv \eta &=
\frac{\xie|_{i}}{\xis|_{i}} \delta\eta_{i}\big[1+O(\delta\eta_{i})\big]~.
\end{align*}
We consolidate the kinetic rates $k_u$, $k_d$ into the 
parameter \(k = 2/(k_{u}^{-1}+k_{d}^{-1})\) of~(\ref{eq:kinetic parameter}). Thus,~(\ref{eq:flux-density 1}) reduces to
\begin{equation}\label{eq:flux-density 3}
\left[\left( \frac{2}{k} +
\frac{\xie|_{i}D_{22}}{|\mathbf{D}^{\ter}|}\delta\eta_{i}\right)
J^{\ter}_{i,\bot}|_{i} -
\frac{\xie|_{i}D_{12}}{|\mathbf{D}^{\ter}|}\delta\eta_{i}
J^{\ter}_{i,\parallel}|_{i}\right]\big[1+O(\delta\eta_{i})\big] = C_{i}^{eq} -
C_{i+1}^{eq}~.
\end{equation}
We multiply~(\ref{eq:flux-density 3}) by
\(|\mathbf{D}^{\ter}|/(\xie|_{i}\delta\eta_{i})\) and thereby obtain 
\begin{equation}\label{eq:flux-density 1b}
\left( D_{22} + \frac{2|\mathbf{D}^{\ter}|}{k\xie|_{i}\delta\eta_{i}}
\right)J^{\ter}_{i,\bot}|_{i} - D_{12} J^{\ter}_{i,\parallel}|_{i} =
|\mathbf{D}^{\ter}| \frac{C_{i}^{eq} -
C_{i+1}^{eq}}{\xie|_{i} \delta\eta_{i}}~.
\end{equation}

As \(\delta\eta_{i} \rightarrow 0,\) the right-hand side of~(\ref{eq:flux-density 1b}) approaches
\(C_{s}|\mathbf{D}^{\ter}|\partial_{\bot}\mu/k_{B}T.\) On the other hand,
the ratio of parameters in the
prefactor of \(J^{\ter}_{i,\bot}|_{i}\) has the limiting value
\begin{equation}\label{eq:new-q def}
\frac{2|\mathbf{D}^{\ter}|}{k\xie|_{i}\delta\eta_{i}} \rightarrow
\frac{2|\mathbf{D}^{\ter}|}{ka}|\nabla h|=\mathcal D^{\ter}\,|\nabla h|~,\qquad \mathcal D^{\ter}:=\frac{2|{\bf D}^{\ter}|}{ka}~,
\end{equation}
where $\mathcal D^{\ter}$ has dimensions of diffusivity [(length)$^2$/time].

A matrix equation for the continuum-scale surface flux \({\bf J}^{\ter}=(J^{\ter}_{\bot},J^{\ter}_{\parallel})^{T}\)
in terms of the step chemical potential \(\mu\) comes from combining~(\ref{eq:flux-density 2}), (\ref{eq:flux-density 1b})
and~(\ref{eq:new-q def}): 
\begin{equation}\label{eq:flux-potential 4}
\left( \begin{array}{cc}
D_{22} + \mathcal D^{\ter}|\nabla h| & \qquad -D_{12} \\
-D_{21} & \qquad \phantom{-}D_{11} \end{array} \right) \left( \begin{array}{c}
J^{\ter}_{\bot} \\
J^{\ter}_{\parallel} \end{array} \right) =
-\frac{C_{s}|\mathbf{D}^{\ter}|}{k_{B}T} \left( \begin{array}{c}
\partial_{\bot}\mu \\
\partial_{\parallel}\mu \end{array} \right)~.
\end{equation}
By solving (\ref{eq:flux-potential 4}) for \(\mathbf{J}^{\ter}\) we obtain
\begin{equation}\label{eq:flux-potential off-diagonal}
\mathbf{J}_i^{\ter}|_i\rightarrow {\bf J}^{\ter}({\bf r},t)=\left( \begin{array}{c}
J^{\ter}_{\bot} \\
J^{\ter}_{\parallel} \end{array} \right) =
-C_{s}\mathbf{M}^{\ter}\cdot\left( \begin{array}{c}
\partial_{\bot}\mu \\
\partial_{\parallel}\mu \end{array} \right)~, 
\end{equation}
where the continuum-scale adatom mobility is
\begin{equation}\label{eq:mobility off-diagonal}
\mathbf{M}^{\ter} = \frac{1}{k_{B}T\left( 1+ q|\nabla
h|\right)} \left( \begin{array}{cc}
D_{11} \qquad & D_{12} \\
D_{21} \qquad & D_{22}+\mathcal D^{\ter}|\nabla h| \end{array} \right)~,\qquad q:=\frac{2D_{11}}{ka}~.
\end{equation}

This formula reduces to the equation with diagonal \(\mathbf{M}^{\ter}\) found
in Ref.~\onlinecite{margetiskohn06} when \(D_{11}=D_{22}=D^{\ter}\) and
\(D_{12}=D_{21}=0\); cf.~(\ref{eq:mobility 0}).  In contrast to the case with scalar diffusivity, all
matrix elements of the mobility in~(\ref{eq:mobility off-diagonal})
depend on the slope. This dependence is quite pronounced in the kinetic regime of
attachment-detachment limited (ADL) kinetics, which we discuss in section
\ref{sec:discussion} below.

\subsection{Alternative approach to continuum: Taylor expansions}
\label{subsec:alter}

For the sake of completeness, we re-derive~(\ref{eq:flux-potential off-diagonal}) and~(\ref{eq:mobility off-diagonal})
via an alternative yet equivalent route. This is based on expansions of the boundary conditions~(\ref{eq:attachment-detachment a})
and~(\ref{eq:attachment-detachment b}) for atom attachment-detachment
in appropriate Taylor series when \(\delta\eta_{i}=\eta_{i+1}-\eta_i\rightarrow 0\) and $\delta\sigma=\sigma'-\sigma\rightarrow 0$.

Following the derivation outlined by one of us in a Letter,~\cite{margetis07} we first expand $C_i|_{i+1}$ and $J_{i,\bot}^{\ter}|_{i+1}$
in~(\ref{eq:attachment-detachment
b}) to first order in \(\delta\sigma\) and \(\delta\eta_{i}\):
\begin{equation}\label{eq:Taylor expansion 2}
k_{u}\left(J^{\ter}_{i,\bot}|_{i} +
\partial_{\eta}J^{\ter}_{i,\bot}|_{i}\delta\eta_{i} +
\partial_{\sigma} J^{\ter}_{i,\bot}|_{i}\delta\sigma\right) =
k_{u}k_{d}\big[C_{i}|_{i}  
+ \partial_{\eta} C_{i}|_{i}\delta\eta_{i}
+ \partial_{\sigma} C_{i}|_{i}\delta\sigma -
C_{i}^{eq}(\sigma+\delta\sigma,t)\big]~.
\end{equation}
Second, we multiply (\ref{eq:attachment-detachment a}) by \(k_{d}\) and
subtract the resulting equation from (\ref{eq:Taylor expansion 2}), so as to eliminate \(C_{i}.\)
By neglecting the \(\eta\)- and \(\sigma\)-derivatives of
\(J^{\ter}_{i,\bot}\), we find 
\begin{equation}\label{eq:density-flux-potential}
(k_{u}+k_{d})J^{\ter}_{i,\bot}|_{i} = k_{u}k_{d}\big\{
\partial_{\eta} C_{i}|_{i}\delta\eta_{i}
+ \partial_{\sigma} C_{i}|_{i}\delta\sigma 
- \frac{C_{s}}{k_{B}T}[\mu(\eta_{i+1},\sigma+\delta\sigma) -
\mu(\eta_{i},\sigma)] \big\}~.
\end{equation}

Next, we solve for \(\partial_{\eta} C_{i}\) and
\(\partial_{\sigma} C_{i}\) by applying the matrix equation
(\ref{eq:density-flux}). The substitution of \(\partial_{\eta} C_{i}\) and \(\partial_{\sigma}
C_{i}\) into~(\ref{eq:density-flux-potential}) and subsequent expansion of the
difference \(\mu(\eta_{i+1},\sigma+\delta\sigma) -\mu(\eta_{i},\sigma)\)
about \((\eta_{i},\sigma)\) yields a relation between
\(\mathbf{J}^{\ter}_{i}\) and the gradient of the continuum step chemical
potential \(\mu(\mathbf{r},t):\)
\begin{multline}\label{eq:flux-potential 2}
\left( \frac{1}{k_{u}} + \frac{1}{k_{d}} + \frac{D_{22}\,\xie
\delta\eta_{i}}{|{\bf D}^{\ter}|} \right) J^{\ter}_{i,\bot}|_{i} -
\frac{\xie D_{12} \delta\eta_{i}}{|{\bf D}^{\ter}|}
J^{\ter}_{i,\parallel}|_{i} + \frac{C_{s}}{k_{B}T}
\partial_{\eta}\mu|_{i}\delta\eta_{i} \\
= \left[ \frac{\xis}{|{\bf D}^{\ter}|} \big(
D_{12}J^{\ter}_{i,\bot} - D_{11}J^{\ter}_{i,\parallel} \big)|_{i} -
\frac{C_{s}}{k_{B}T}
\partial_{\sigma}\mu|_{i} \right]\delta\sigma~.
\end{multline}

Setting \(\delta\sigma=0\) in (\ref{eq:flux-potential 2}) and taking
the continuum limit provides our first equation for the components of
the surface flux in terms of \(\mu:\)
\begin{equation}\label{eq:transverse mu}
\left( 1+ \frac{2|\mathbf{D}^{\ter}|}{kaD_{22}}|\nabla h| \right)J^{\ter}_{\bot} -
\frac{D_{12}}{D_{22}}J^{\ter}_{\parallel} =
-\frac{C_{s}|\mathbf{D}^{\ter}|}{k_{B}TD_{22}} \partial_{\bot}\mu~.
\end{equation}

The continuum limit of (\ref{eq:flux-potential 2}) still applies when
\(\delta\sigma \neq 0.\) By~(\ref{eq:transverse mu}), we know that
the left-hand side of (\ref{eq:flux-potential 2}) tends to zero in that
limit.  Therefore, the term proportional to \(\delta\sigma\)
must also vanish as \(\delta\eta_{i}\rightarrow 0\). Thus, we have
\begin{equation}\label{eq:parallel mu}
D_{21}J^{\ter}_{\bot} - D_{11}J^{\ter}_{\parallel} =
\frac{C_{s}|\mathbf{D}^{\ter}|}{k_{B}T}\partial_{\parallel}\mu~.
\end{equation}

By solving simultaneously~(\ref{eq:transverse mu}) and~(\ref{eq:parallel
mu}) for the components of the continuum surface flux, we find
\begin{equation}\label{eq:continuum flux}
\left( \begin{array}{c}
J^{\ter}_{\bot} \\
J^{\ter}_{\parallel} \end{array}\right) = \frac{-C_{s}}{k_{B}T\left( 1+
q|\nabla{} h| \right)} \left( \begin{array}{cc}
D_{11} \qquad & D_{12} \\
D_{21} \qquad & D_{22} + \mathcal D^{\ter}\, |\nabla{} h| \end{array}
\right)\cdot\left( \begin{array}{c}
\partial_{\bot}\mu \\
\partial_{\parallel}\mu \end{array} \right)~,\qquad \mathcal D^{\ter}=\frac{2|{\bf D}^{\ter}|}{ka}~,\quad q=\frac{2D_{11}}{ka}~,
\end{equation}
which is directly identified with the combination of~(\ref{eq:flux-potential
off-diagonal}) and~(\ref{eq:mobility off-diagonal}).

\subsection{Mass conservation law and total surface flux}
\label{subsec:mass conservation}

In this subsection we define the {\it total} surface flux ${\bf J}$ so that the
mass conservation law for atoms is satisfied in the presence of step edge diffusion. 
The surface mobility is defined accordingly through the relation
of ${\bf J}$ and $\mu$.

At a given location \(\sigma\) on the \(i\)th step edge, the step normal
velocity \(v_{i}\) must respect conservation of mass, taking into
account all possible sources and sinks of atoms; see~(\ref{eq:step velocity}).  
By the discussion of Sec.~\ref{sssec:mass-conserv}, in the continuum limit~(\ref{eq:step velocity}) reduces to
\begin{equation}\label{eq:step velocity continuum}
\partial_{t}h = -\Omega\nabla\cdot\mathbf{J}^{\ter} + \frac{a|\nabla
h|}{\xis}\partial_{\sigma}\left\{\frac{D^{\edge}}{\xis}\partial_{\sigma}
\left(\frac{\mu}{k_{B}T} \right)\right\}~,
\end{equation}
where the adatom flux ${\bf J}^{\ter}$ is described by~(\ref{eq:flux-potential
off-diagonal}) and~(\ref{eq:mobility off-diagonal}).

Since the terrace is a level set for the height, we have
\(h=H(\eta,t)\); in other words, \(h\) does not vary in the
step-longitudinal ($\sigma$-) direction.  Thus,
\(|\nabla h| = \xie^{-1}|\partial_{\eta}H|\) and the factor
\(|\partial_{\eta}H|\) can be passed through the \(\sigma\) derivative
in (\ref{eq:step velocity continuum}).  It follows that
\begin{equation}\label{eq:continuum step velocity 0}
\partial_{t}h = -\Omega\nabla\cdot\mathbf{J}^{\ter} + \frac{1}{\xie\xis}
\partial_{\sigma}\left\{aD^{\edge}|\nabla h|
\frac{\xie}{\xis}\partial_{\sigma} \left(\frac{\mu}{k_{B}T}
\right)\right\}~.
\end{equation}
We recognize the second term on the right-hand side of
(\ref{eq:continuum step velocity 0}) as the divergence of 
\(a D^{\edge}|\nabla h| \partial_{\parallel}(\mu/k_{B}T) \longv.\)  Hence, we
refer to the term \(-\frac{a D^{\edge}}{\Omega}|\nabla h|
\partial_{\parallel}(\mu/k_{B}T) \longv\) as the edge atom flux, denoted by
\(\mathbf{J}^{\edge}.\) Combining the two divergence terms into one term, we
obtain the mass conservation law
\begin{equation}\label{eq:mass conservation 0}
\partial_{t}h = -\Omega\nabla\cdot(\mathbf{J}^{\ter} +
\mathbf{J}^{\edge}) = -\Omega\nabla\cdot\mathbf{J}~,
\end{equation}
where 
\begin{equation}\label{eq:edge flux}
{\bf J}={\bf J}^{\ter}+{\bf J}^{\edge}~,\qquad\mathbf{J}^{\edge}:=-\frac{aD^{\edge}}{\Omega}|\nabla h|\,\partial_\parallel
\biggl(\frac{\mu}{k_BT}\biggr)\longv~.
\end{equation}

Thus, the matrix equation~(\ref{eq:continuum flux}) involving the mobility tensor can be updated
accordingly for the effective surface flux:
\begin{equation}\label{eq:updated_flux}
{\bf J}({\bf r},t)=\left( \begin{array}{c}
J_{\bot} \\
J_{\parallel} \end{array} \right) =
-C_{s}\left( \begin{array}{cc}
M_{\eta\eta} & M_{\eta\sigma} \\
M_{\sigma\eta} & M_{\sigma\sigma} \end{array} \right)
\cdot \left( \begin{array}{c}
\partial_{\bot}\mu \\
\partial_{\parallel}\mu \end{array} \right)=-C_s{\bf M}\cdot \nabla\mu~, 
\end{equation}
where
\begin{equation}\label{eq: mobility-matr}
{\bf M}=\left(\begin{array}{cc} M_{\eta\eta} & M_{\eta\sigma}\\
                                M_{\sigma\eta} & M_{\sigma\sigma}
              \end{array}\right)~,
\end{equation} 
\begin{align}\label{eq:M_updated_elements}
M_{\eta\eta} &= \frac{D_{11}}{k_BT}\,\frac{1}{\displaystyle 1+2\frac{D_{11}}{ka}|\nabla
h|}~, \qquad
M_{\eta\sigma} = \frac{D_{12}}{k_BT}\,\frac{1}{\displaystyle
1+2\frac{D_{11}}{ka}|\nabla h|}~, \nonumber \\
M_{\sigma\eta} &= \frac{D_{21}}{k_BT}\,\frac{1}{\displaystyle
1+2\frac{D_{11}}{ka}|\nabla h|}~, \qquad
M_{\sigma\sigma} = \frac{1}{k_BT}\,\frac{\displaystyle D_{22} +
\frac{2|\mathbf{D}^{\ter}|}{ka}|\nabla h|}{\displaystyle
1+2\frac{D_{11}}{ka}|\nabla h|} +\frac{aD^{\edge}}{\Omega C_{s}}|\nabla h|~.
\end{align}

In applications it is often desirable to represent the total mobility tensor ${\bf M}$ with respect to
a fixed coordinate system.  We invoke the similarity transformation
outlined in Ref.~\onlinecite{margetiskohn06} in order to obtain the basal plane's Cartesian representation
of \(\mathbf{M}\).  Using the change-of-basis matrix
\begin{equation}\label{eq:S def}
\mathbf{S} = |\nabla h|^{-1} \begin{pmatrix} -\partial_{x}h &
\partial_{y}h \\
-\partial_{y}h & -\partial_{x}h \end{pmatrix}~,
\end{equation}
we obtain the representation
\begin{equation}
\mathbf{M}_{(x,y)} = \mathbf{S}\,\mathbf{M}\,\mathbf{S}^{-1}=\frac{\widetilde M_{xx}\ux\ux + \widetilde M_{xy}\ux\uy +
\widetilde M_{yx}\uy\ux + \widetilde M_{yy}\uy\uy}{\displaystyle{k_{B}T|\nabla h|^{2} \left(
1+\frac{2D_{11}}{ka}|\nabla h|\right)}}~,
\end{equation}
where
\begin{align}
\widetilde M_{xx} &:= D_{11}(\partial_x h)^2-(D_{12}+D_{21})(\partial_xh)(\partial_y h)+\biggl[\big(D_{22}+\mathcal D^{\ter}
|\nabla h|\big)+\frac{a D^{\edge}}{\Omega C_s}|\nabla h|\biggl(1+\frac{2D_{11}}{ka}|\nabla h|\biggr)\biggr](\partial_y h)^2~,\label{eq:Mxx}\\
\vspace*{2em} \widetilde M_{xy} &:= D_{12}(\partial_x h)^2-D_{21}(\partial_y h)^2+\biggl[D_{11}-\big(D_{22}+\mathcal D^{\ter}|\nabla h|\big)-
\frac{aD^{\edge}}{\Omega C_s}|\nabla h|\biggl(1+\frac{2D_{11}}{ka}\biggr)\biggr](\partial_x h)(\partial_y h)~,\label{eq:Mxy}\\
\vspace*{2em} \widetilde M_{yx} &= D_{21}(\partial_x h)^2-D_{12}(\partial_y h)^2+\biggl[D_{11}-\big(D_{22}+\mathcal D^{\ter}|\nabla h|\big)-
\frac{aD^{\edge}}{\Omega C_s}|\nabla h|\biggl(1+\frac{2D_{11}}{ka}\biggr)\biggr](\partial_x h)(\partial_y h)~,\label{eq:Myx}\\
\vspace*{2em} \widetilde M_{yy} &= \biggr[\big(D_{22}+\mathcal D^{\ter}\,|\nabla h|\big)+\frac{a D^{\edge}}{\Omega C_s}|\nabla h|
\biggl(1+\frac{2D_{11}}{ka}|\nabla h|\biggr)\biggr](\partial_x h)^2+(D_{12}+D_{21})(\partial_x h)(\partial_y h)+D_{11}(\partial_y h)^2~.\label{eq:Myy}
\end{align}

So far, we derived a relation of the form ${\bf J}=-C_s{\bf M}\cdot\nabla \mu$ for the surface flux
where $\partial_t h=-\Omega\,{\rm div}{\bf J}$. The chemical potential $\mu$ is related to derivatives
of $h$ through~(\ref{eq:continuum mu}).

\subsection{PDE for height profile}
\label{subsec:height PDE}

We now combine the mass conservation law~(\ref{eq:mass conservation 0})
with the effective surface flux~(\ref{eq:updated_flux}) and the formula for the continuum step chemical potential~(\ref{eq:continuum mu})
in order to derive a PDE analogous to~(\ref{eq:PDE-2D}) for the surface height
profile, \(h(\mathbf{r},t)\).  With the substitutions for $\mu$ and \(\mathbf{J}\)
by~(\ref{eq:continuum mu}) and~(\ref{eq:updated_flux}), the mass conservation law~(\ref{eq:mass conservation 0}) becomes
\begin{equation}\label{eq:height evolution}
\partial_{t}h = -\frac{\Omega^2 C_s}{a}
{\rm div}\left\{\mathbf{M}\cdot\nabla\left(
{\rm div}\left[(\beta + \tilde g|\nabla h|^{2})\frac{\nabla
h}{|\nabla h|}\right] \right) \right\}~.
\end{equation}
To consolidate the physical parameters, we define \(g_{1}=\beta/a,\)
\(g_{3} = \tilde{g}/a,\) and \(B=\Omega^{2}C_{s}g_{1}\); see~(\ref{eq:parameters}).
Accordingly, we obtain~(\ref{eq:PDE-2D}) with ${\bf M}^{\ter}$ replaced by the effective total mobility ${\bf M}$.

%
%
\section{Scaling laws}
\label{sec:discussion}

In this section we derive approximate, separable solutions of PDE~(\ref{eq:height evolution}). Our goal
is to find plausible connections of actual continuum solutions to decay laws observed in biperiodic profiles,
e.g. observations reported in Refs.~\onlinecite{keefeetal94,blakelyetal97,erlebacher00,pedemonteetal03}. Our discussion
is heuristic; the relation of PDE solutions to experiments is not
well understood at the moment.

We start with the ansatz \(h({\bf r},t)\approx A(t)H(\mathbf{r})\). This
separation of variables, called a ``scaling ansatz'', is consistent with 
previously reported step flow simulations in 1D~\cite{israelikandel00} and
kinetic Monte Carlo simulations in 2D,~\cite{shenoy04} both with initial
sinusoidal profiles. The amplitude $A(t)$ can be obtained formally from an ordinary differential equation (ODE)
by direct substitution in~(\ref{eq:height evolution}). We alert the reader that 
conditions on the initial data and material parameters for having separable solutions and recovering an ODE for $A$ 
are currently elusive, requiring detailed numerical studies. Such studies
lie beyond our present scope. 

Additive terms in the driving force \(\nabla \mu\) and in the total
mobility \(\mathbf{M}\) scale differently with \(A.\) We need to retain in the right-hand side of the PDE terms
proportional to the same power of \(A\) and thus resort to approximations. It should be borne in mind that the nonlinearities 
in ${\bf M}$ and $\mu$ lead to spatial-frequency coupling for biperiodic height profiles; accordingly,
evolution is in principle more complicated than the one implied here
by our simple scaling scenario.

Depending on the powers of \(A\) that possibly prevail in the evolution equation,
we find several plausible behaviors of $h$ with time, including the exponential decay and
inverse linear decay reported in related experiments.~\cite{keefeetal94,blakelyetal97,erlebacher00,pedemonteetal03}  
By~(\ref{eq:continuum mu}) the driving
force \(\nabla\mu\) scales as \(A^{0}\) if the dominant term is step
line tension. If step interactions are
dominant, then \(\nabla\mu\) scales as
\(A^{2}.\)  To determine the scaling of the mobility tensor, it is
convenient to introduce the ``aspect ratio" \(\alpha :=
\partial_{y}h/\partial_{x}h\); it is plausible yet not compelling to estimate $\alpha$ by  $\lambda_x/\lambda_y$ 
where $\lambda_x$ and $\lambda_y$ are wavelengths in the $x$ and $y$ directions. We also define the
slope-dependent quantity \(b := (1+\frac{2D_{11}}{ka}|\nabla
h|)^{-1}.\)  Note that $\alpha$ scales as $A^0$. When step edge diffusion is absent ($D^{\edge}=0$),
the possible scalings found for $A$ with nonzero $D_{12}$ and $D_{21}$ are not different from those for isotropic adatom diffusion
(where $D_{12}=D_{21}=0$).~\cite{margetis07}

With these definitions, the elements $M_{ij}=(k_BT)^{-1}|\nabla h|^{-2}b\widetilde M_{ij}$ 
($i,\,j=x,\,y$) from the Cartesian representation~(\ref{eq:Mxx})--(\ref{eq:Myy}) of ${\bf M}$ read
\begin{align}
M_{xx} &= \frac{b\,(\partial_{x}h)^{2}}{k_{B}T|\nabla h|^{2}} \left[
D_{11} - \alpha(D_{12}+D_{21}) + \alpha^{2}D_{22} +
\frac{2|\mathbf{D}^{\ter}|}{ka}
\alpha^{2}|\nabla h| + \frac{aD^{\edge}\alpha^{2}|\nabla h|}{b\,\Omega
C_{s}} \right]~, \nonumber \\
M_{xy} &= \frac{b\,(\partial_{x}h)^{2}}{k_{B}T|\nabla h|^{2}} \left[
D_{12} + \alpha(D_{11} - D_{22}) - \alpha^{2}D_{21} -
\frac{2|\mathbf{D}^{\ter}|}{ka}\alpha|\nabla h| - \frac{aD^{\edge}\alpha|\nabla
h|}{b\,\Omega C_{s}} \right]~, \nonumber \\
M_{yx} &= \frac{b\,(\partial_{x}h)^{2}}{k_{B}T|\nabla h|^{2}} \left[
D_{21} + \alpha(D_{11} - D_{22}) - \alpha^{2}D_{12} -
\frac{2|\mathbf{D}^{\ter}|}{ka}\alpha
|\nabla h| - \frac{aD^{\edge}\alpha|\nabla h|}{b\,\Omega C_{s}}
\right]~, \nonumber \\
\label{eq:Tmobility elements}
M_{yy} &= \frac{b\,(\partial_{x}h)^{2}}{k_{B}T|\nabla h|^{2}} \left[
D_{22} + \alpha(D_{12} + D_{21}) + \alpha^{2}D_{11} +
\frac{2|\mathbf{D}^{\ter}|}{ka}|\nabla
h| + \frac{aD^{\edge}|\nabla h|}{b\,\Omega C_{s}} \right]~.
\end{align}

We restrict attention to ADL kinetics which closely correspond to relevant
experimental situations.~\cite{keefeetal94,blakelyetal97,erlebacher00,pedemonteetal03} 
It follows that \(b \ll 1\) where \(b\) scales
as \(A^{-1};\)  by the scaling ansatz for $h$, the
prefactor \(\frac{b(\partial_{x}h)^{2}}{k_{B}T|\nabla h|^{2}}\)
also scales as \(A^{-1}.\)  For the sake of simplicity we consider 
weak anisotropy, \( |\mathbf{D}^{\ter}| \approx D_{11}D_{22} \)
(i.e., if the off-diagonal diffusivity elements \(D_{12}, D_{21}\) are small
in comparison to the diagonal elements) and
\(|\mathbf{D}^{\ter}|/(ka) \gg aD^{\edge}/(b\,\Omega
C_{s})\). The dominant terms in ${\bf M}$ scale as:

(i) $A^0$ if $b\ll \min\{(D_{22}/D_{11})\alpha^2,(D_{22}/D_{11})\alpha^{-2},D_{22}/D_{11}\}$; and 

(ii) $A^{-1}$ if $b\gg \max\{(D_{22}/D_{11})\alpha^2,(D_{22}/D_{11})\alpha^{-2},D_{22}/D_{11}\}$. 

\noindent In presence of step edge diffusion with \( |\mathbf{D}^{\ter}|/(ka) \ll aD^{\edge}/(b\,\Omega
C_{s}),\) the dominant terms in the mobility tensor scale as \(A^{1}.\)
Note that in all theses cases the matrix ${\bf M}$ tends to become singular since the lowest
eigenvalue acquires a small value. Hence, correction terms in ${\bf M}$, which strictly spoil the 
scalings reported here, are physically important; solutions of the form $A(t) H({\bf r})$
should be thought of as leading-order terms of appropriate asymptotic expansions for $h$.\looseness=-1

Next, we combine the three possible scalings of
\(\mathbf{M}\) with the two possible scalings of
\(\nabla\mu.\) Each combination yields an ODE of the form \(\dot{A} \propto -A^{p}\) for
some exponent \(p;\) the minus sign here is assumed for achieving profile decay. 
In the case of ADL kinetics, outlined above, we
have \(p \in \{ -1,0,1\} \cup \{1,2,3\},\) where the first set
corresponds to dominant step line tension and
the second set corresponds to dominant step interactions in $\nabla\mu$.  Since \(p=1\) is common to both sets, the associated
scaling law \(A = A_{0}\exp(-t/\tau)\) could perhaps be observed in a wide range of
experimental situations.  On the other hand, the scaling law \(A =
A_{0}/\sqrt{1+t/\tau}\) associated with \(p=3\) and dominance of step edge diffusion may not be physical;
to our knowledge, this last decay law has not been observed.

We illustrate the procedure of finding \(A\) for weak anisotropy under condition (ii) above
and dominant step interactions; thus, $p=1$. The PDE becomes
\begin{equation}\label{eq:sep-H}
\dot{A}(t)H(\mathbf{r}) = -\frac{\Omega^{2} C_{s} g_{3}}{k_{B}T}
\frac{ka\,A(t)}{2D_{11}}\ {\rm div}\Biggl\{\frac{(\partial_x H)^2}{|\nabla H|^3}\,
\begin{pmatrix}
m_{xx} & m_{xy} \\
m_{yx} & m_{yy}
\end{pmatrix}\cdot\nabla\big[{\rm div}\big(|\nabla H|\nabla H\big)\big]\Biggr\}~,
\end{equation}
where the elements \(\{m_{ij}\}_{i,j=x}^{y}\) are constants that stem from ${\bf M}_{(x,y)}$ after
factoring out \(A\) (but not \(H\)); the precise definition of $m_{ij}$ is omitted here.

To satisfy~(\ref{eq:sep-H}) for all \(t\) and \(\mathbf{r},\) we require
that the time-dependent part \(A(t)\) solve \(\dot{A}(t) =
-\mathcal{C}A\) for some positive constant \(\mathcal{C}\) ($\mathcal C>0$).  The height profile
\(H(\mathbf{r})\) solves the nonlinear PDE
\begin{equation}\label{eq:PDE for H}
\mathcal C\,H=\frac{\Omega^{2} C_{s} g_{3}}{k_{B}T}
\frac{ka}{2D_{11}}\ {\rm div}\Biggl\{\frac{(\partial_x H)^2}{|\nabla H|^3}\,
\begin{pmatrix}
m_{xx} & m_{xy} \\
m_{yx} & m_{yy}
\end{pmatrix}\cdot\nabla\big[{\rm div}\big(|\nabla H|\nabla H\big)\big]\Biggr\}~.
\end{equation}
The solution for \(A(t)\) is given in terms of the separation constant
\(\mathcal{C}\) and the initial amplitude \(A_{0}\): $A(t) = A_{0}e^{-\mathcal{C}t}$.
Using a similar procedure, we derive other possible scaling laws
for ADL kinetics under different restrictions.  Our results are summarized in Table~\ref{T:scaling laws}.

We do not address the issue of solving~(\ref{eq:PDE for H}) in this
analysis.  Particularly interesting is the case with facets. The continuum limit breaks down at facet edges and 
associated boundary conditions for \(H\) must take into account the
discrete step flow equations.~\cite{margetisetal06}  A numerical scheme to implement these boundary
conditions within continuum is still under development.  

\begin{table}\caption{\label{T:scaling laws} 
Decay laws for
height amplitude \(A(t)\) in ADL kinetics. Leftmost column indicates plausible
conditions. Next two columns list respective decay laws for line tension and step interaction dominated $\nabla\mu$.
The time constant \(\tau\) depends on \(A(0)\) and \(H.\)}
\begin{tabular}{c | c c} \hline
& Line tension & Step interaction \\ \hline
$|{\bf D}^{\ter}|\approx D_{11}D_{22}$ & & \\
\( b\gg \max\{(D_{22}/D_{11})\alpha^2,D_{22}/D_{11},(D_{22}/D_{11})\alpha^{-2}\} \) & \( A_{0}\sqrt{1-t/\tau} \) & \(
A_{0}\exp(-t/\tau) \) \\
\( b\ll \min\{D_{22}/D_{11})\alpha^2,D_{22}/D_{11},(D_{22}/D_{11})\alpha^{-2}\}  \) & \(A_{0}(1-t/\tau)\) & \(A_{0}/(1+t/\tau)\) \\ \hline
\( |\mathbf{D}^{\ter}|/(ka) \ll aD^{\edge}/(b\Omega
C_{s}) \)
& \( A_{0}\exp(-t/\tau)\) & \(A_{0}/\sqrt{1+t/\tau} \) \\ \hline
\end{tabular}
\end{table}

A similar analysis can be carried out if terrace diffusion is the
slowest process, i.e., \(q|\nabla h|=|\nabla h|D_{11}/(ka) \ll 1.\)  Then, \(b\)
is approximately a constant, \(b \approx 1.\) The dominant terms in the mobility tensor scale as
\(A^{0}\) or \(A^{1}\).  Thus, we obtain \(\dot{A}\propto -A^{p}\) for
\(p\in\{0,1,2,3\},\) which yields four of the five decay laws already found for ADL kinetics.

\section{Conclusion}
\label{sec:conclusion}

By interpreting a (2+1)-dimensional step flow model for a relaxing surface as a
discretization of a continuum evolution equation, we derived the relevant PDE for the surface height profile.
The starting point is a step velocity law that accounts for anisotropic adatom diffusion
on terraces, diffusion of atoms along step edges and atom attachment-detachment at steps. 
In the continuum limit we obtained a relation between the surface flux
and the step chemical potential. This relation involves a tensor surface mobility as an effective 
coefficient. 

We gave two different derivations of the surface mobility under
the assumption of linear kinetics at step edges.  Our main approach relies
on the direct solution of the diffusion equation for adatoms on
each terrace via the separation of local step coordinates into fast and slow.
The continuum limit is attained by letting the
step height and terrace widths tend to zero under the condition that
the slope remains finite. 

Combining the step velocity law with the continuum relation
between the surface flux and the step chemical potential resulted in a
nonlinear, fourth-order parabolic PDE for the surface height.  Transforming the
mobility tensor from local step coordinates to fixed coordinates
induced a dependence on the height partial derivatives.  This
dependence offers a plausible
scenario of how an epitaxial surface can exhibit different decay laws.
We found separable solutions for the height that approximately satisfy the 
evolution equation under certain conditions.  These separable solutions exhibit
different decay and may be used as a guide in interpreting experimental observations
from a continuum viewpoint.

Our PDE only accounts for a part of the possible microscopic physics.  We
neglected elasticity which may induce long-range interactions between steps, surface
reconstruction, material deposition, and evaporation/condensation (sublimation).  
Incorporating these processes into the theory is work in progress.  For example, the
inclusion of evaporation/condensation requires only an additive
term in the step velocity law.~\cite{spohn93} The continuum limit with
this additional effect is already within the scope of the analysis
presented here.  More challenging is the inclusion of processes
that modify: (i) the terrace diffusion equation; (ii) the kinetic boundary
conditions at step edges; and (iii) the formula for the step chemical potential.

The tensor mobility depends crucially on the kinetics of each terrace.  More general mobility
tensors might emerge by encompassing terms that account for (i)--(iii) above.  
With the inclusion of
step edge diffusion, which was absent from previous derivations of a tensor
mobility,~\cite{margetiskohn06,margetis07} we found an
effective mobility whose elements still depend only on \(|\nabla h|\);
even then, the mobility ${\bf M}$ does not involve powers of \(|\nabla h|\)
greater than 1.  We plan to investigate the possible
structure of \(\mathbf{M}\) in more general physical settings. 

The PDE we derived for the surface height may admit separable solutions
under certain conditions, which are not precisely known at the moment. We hope to make
connections to experiments on surface relaxation with anisotropic diffusivity.
One challenge in making these comparisons is to single out experimentally
measurable quantities that correspond to PDE solutions in an appropriate sense.  
Another challenge in this context is the incorporation of facets within a viable
scheme of solving the PDE. The theory
presented here can serve as a basis for future work, in which the PDE for
surface height evolution is implemented numerically for comparisons with
experimental data.

\section*{ACKNOWLEDGMENTS}

We are indebted to Theodore L. Einstein, Joachim Krug, Ray J. Phaneuf, and Ellen D. Williams
for useful discussions.

\appendix

\end{document}